\documentstyle[12pt]{article}
\begin{document}

\begin{quote}

\large
\centering {\bf {Dynamic Susceptibility and Phonon Anomalies \\
in the Bilayer $t$-$J$ Model}}

\bigskip\smallskip
\normalsize

B. Normand, H. Kohno and H. Fukuyama

\bigskip

Department of Physics, University of Tokyo, \\
7-3-1 Hongo, Bunkyo-ku, Tokyo 113, Japan.

\end{quote}

	We consider a bilayer version of the extended $t$-$J$ model, with
a view to computing the form of certain experimentally observable properties.
Using the slave-boson decomposition, we show at the mean-field level that
in the bilayer system the existence of in-plane $d$-wave singlet pairing
excludes any interplane singlet order for reasonable values of the interplane
superexchange parameter. Restricting the analysis to the regime of no
interplane singlet pairing, we deduce parameter sets reproducing the Fermi
surfaces of YBCO- and BSCCO-like bilayer systems. From these we calculate
the form of the dynamic susceptibility $\chi( {\bf q}, \omega )$ in both
systems, and of the anomalies in frequency and linewidth of selected
phonon modes in YBCO. We compare the results with experiment, and discuss
the features which differ from the single-layer case.

\bigskip
\noindent
PACS numbers: 74.20.Mn, 74.25.Ha, 74.25.Kc

\newpage

\section{Introduction}

	Bilayer models for high-$T_c$ superconductors have become
popular recently, not only due to the obvious feature that the
best-characterised material has a bilayer structural unit, but also
since the suggestion by Millis and Monien \cite{rmm1} that the
single- and bilayer systems may have fundamentally different
magnetic excitations. These differences may include the existence
or absence of the spin gap \cite{ry,rr}, which has become a crucial
experimental feature and theoretical benchmark.

	In this work we will pursue an approach based on the $t$-$J$
model \cite{ra,rzr} analysed by the slave-boson decomposition
\cite{rbza,rgjr,rkl,rSHF}. In the mean-field
approximation, this approach has been studied intensively
\cite{rTKuF,rTKFs,rszllk,rlw,rls,rmw,rul,rTKF} with a view to
describing the magnetic properties of the high-$T_c$ materials. As an
example of such investigations, the extended $t$-$J$ model of a single
CuO$_2$ layer, has been shown \cite{rTKFs,rTKF} to give a good account
at the RPA level of many features of the spin excitations in the
monolayer La$_{2-x}$Sr$_x$CuO$_4$ (LSCO) system, and some of those in
the bilayer YBa$_2$Cu$_3$O$_{7-\delta}$ (YBCO) case. These authors
considered the (in)commensuration of magnetic fluctuations, the
frequency dependence of the dynamic susceptibility and the different
temperature dependences of the shift and rate of nuclear magnetic
resonance between high- and low-doping regions. Such models also
contain a consistent description of the ``spin-gap'' phenomenon
\cite{rr}, first noted by Yasuoka \cite{ry}. Taking fluctuations
about the mean-field solution into account by a gauge-field approach
\cite{rnlln}, it is possible to gain an understanding of various
transport properties, including the temperature dependence of the
resistivity, thermopower and Hall coefficient. Again within the same
framework, it has recently been shown \cite{nkf} that one may obtain
a good account of several lattice-related features of the electronic
system, such as phonon anomalies and an isotope effect, by
considering the coupling which arises naturally between phonon modes
of the layer and the spin sector.

	In view of this degree of correspondence to experiment, we
consider the extended $t$-$J$ model to be one of the leading
candidates for a framework in which to construct a coherent
understanding of the many and complex features of the high-$T_c$
problem, and thus that the detailed computation of physical quantities
within it is a valuable exercise. Here we wish to apply the ideas of
the single-layer studies \cite{rTKF} and \cite{nkf} to a consistent
bilayer model, and thus to elucidate the successes and limitations of
the mean-field approach to spin-dependent, microscopic properties.
We draw attention to a brief study \cite{rlw} which established that
the bilayer dynamic susceptibility has the experimentally observed
$q_z$ periodicity. Ubbens and Lee \cite{rul} have also considered
static magnetic properties in a bilayer, nearest-neighbour $t$-$J$
model, including gauge-field effects, and argue for an anisotropic
(or ``extended'') $s$-wave pairing state of coupled intra-and
interplane singlet order, as well as for enhanced spin-gap formation.
Similar results for the spin gap were obtained in Ref. \cite{rilma}.

	The question of the gap
symmetry has remained a topic of much debate: while it is known in
the single-layer model that the low-energy state is $d$-symmetric
\cite{rgjr,rkl}, a variety of reports find somewhat different behaviour
in the presence of coupling to a second layer. In addition to the
result of Ref. \cite{rul}, the weak-coupling antiferromagnetic spin
fluctuation theory, which favours $d$-symmetry in a single plane, has
also been shown to favour extended $s$-symmetry in a bilayer
\cite{rlma}. A recent numerical analysis by Eder {\it {et al.}}
\cite{reom} gives the evolution of the in-plane $d$-wave state as
interplane coupling is increased, and demonstrates in a small cluster
that there is an abrupt crossover to a predominantly interplane
ordered state at some value of the coupling. We will discuss these
results in comparison with our own in section 2.

	The bilayer spin system has also been studied extensively by
Millis and coworkers \cite{rmm1,rilma,rmm2,rmim}, with the aim of
elucidating the quantum nature of the ground state and explaining the
contrasting NMR results in single- and bilayer materials. From
spin-wave and scaling theories \cite{rmm2} it has been shown that
interlayer coupling acts to provide a strong enhancement of the spin
gap state (by suppression of competing magnetic order), which provides
an explanation of the differences without the requirement that the
single- and double-layer systems have completely different ground
states. These
considerations have been cast in a more microscopic form to describe
electrons on a bilayer with a weak interplane antiferromagnetic
interaction \cite{rilma}: the system may vary between a Fermi liquid
state, which is proposed to describe the doping regime of materials
with no spin gap, and a spin liquid picture of fermions interacting
via a gauge field, which shows spin-gap features.

	The structure of this paper is as follows. In section
2 we consider the mean-field equations of the
bilayer problem, deduce the nature of the gap and solve for BSCCO-
and YBCO-like systems. In section 3 we calculate the dynamic
susceptibility for both types of system, and illustrate the
dependence of the results on the wavevector component $q_z$. In
section 4 we compute the superconductive anomalies in $c$-axis
phonon modes of in-plane oxygen atoms for the YBCO system, showing
the contrast between modes of even ($g$) and odd ($u$) symmetry.
Section 5 contains a summary and concluding discussion.

\section{Mean-Field Solution}

	We consider a three-dimensional system consisting of a
periodic but uncoupled stack of coupled CuO$_2$ bilayers, in which
the intra- and interplane hopping ($t$) and superexchange ($J$)
interactions are as shown in Fig.~1.
Denoting the $c$-axis dimension of the unit cell by $c$ and the
separation of the planes of the bilayer by $d$, the ratio $r_c =
d / c$ will appear in a phase factor arising from the bilayer
spacing. The Hamiltonian for a system of $N_z$ bilayer units is
\begin{eqnarray}
H & = & \sum_{n = 1}^{N_z} \left\{ \sum_{l = 1,2} \left[ - \sum_{ij
\sigma} t_{ij} a_{i \sigma}^{l \dag} a_{j \sigma}^{l} + \sum_{\langle
ij \rangle } J_{ij} {\bf S}_{i}^{l} {\bf {.S}}_{j}^{l} \right]
\right. \label{eh} \nonumber \\ & & \;\;\;\;\;\; \left. - \sum_{ij
\sigma} t_{\perp ij} \left( a_{i \sigma}^{1 \dag} a_{j \sigma}^{2}
+ a_{i \sigma}^{2 \dag} a_{j \sigma}^{1} \right) + \sum_{ i }
J_{\perp} {\bf S}_{i}^{1} {\bf {.S}}_{i}^{2}
\right\} ,
\end{eqnarray}
where $l$ is the layer index in each unit cell. The first line
contains the $t$-$J$ model of a single CuO$_2$ layer, where
$t_{ij}$ denotes the extended transfer integrals, $J_{ij}$ the
superexchange interaction, which is assumed to be finite only
between nearest neighbours, and the Hilbert space excludes double
occupancy of the quasiparticles $a_{i \sigma}$. The second line
contains the coupling between the layers, which may be by both
hopping and superexchange. Within the plane, the nearest-neighbour
hopping term is taken to be $t = 4 J$, while second- and
third-neighbour terms will be chosen below in order to match the
Fermi surface shape of the physical systems \cite{rTKFs}. We will
consider initially a range of values for the interlayer coupling
parameters $t_{\perp ij}$ and $J_{\perp}$.

	The assumption of coherent, single-particle hopping
processes between the planes of the bilayer is counter to the
interlayer pair tunnelling theories \cite{raipt}, which demands that
this vanish by orthogonality. A clear experimental indication of
whether interlayer transfer is coherent or incoherent should be given
by photoemission experiments, which will see either two Fermi surfaces
or one. At the moment there is some controversy, and inconsistency,
surrounding the interpretation of results obtained for both BSCCO
\cite{rsg,rag} and YBCO \cite{ra124,ra123} systems, part of which
centres on the issue of the number of true band-crossings observed
where the bonding and antibonding bands of the coherently-coupled
bilayer are expected to be far apart in the Brillouin zone. Here we
avoid further discussion of this topic and proceed to elucidate the
properties of the bilayer system with coherent hopping, but will
refer below to the 2-band interpretations of the experimental data.

	In the slave-boson decomposition, the operator $a_{j s}$
is represented as $a_{j s} = f_{j s} b_{j}^{\dag}$, where $f_{j s}$
is a fermion (spinon) carrying the spin degrees of freedom
${\bf S}_j$, and $b_j$ a bosonic holon carrying the charge. In the
mean-field approximation, the spin and charge behave independently,
and all 4-operator terms may be written expressions quadratic in the
operators with the mean-field order parameters (finite
expectation values of possible 2-operator combinations) as coefficient,
while terms quadratic in these order parameters are also generated.
Here we take as possible in-plane order parameters

\smallskip

i) $\chi_{i,i+\tau} = \sum_{\sigma} \langle f_{i \sigma}^{l \dag}
f_{i+\tau \sigma}^{l} \rangle$: uniform RVB of spinons

\vskip2mm

ii) $\Delta_{i,i+\tau} = \mbox{$\frac{1}{\sqrt{2}}$} \langle f_{i
\uparrow}^{l} f_{i+\tau \downarrow}^{l} - f_{i \downarrow}^{l}
f_{i+\tau \uparrow}^{l} \rangle$: singlet RVB

iii) $B_{i,i+\tau} = \langle b_{i \sigma}^{l} b_{i+\tau \sigma}^{l
\dag} \rangle$: uniform RVB of holons

\smallskip
\noindent
and as interplane order parameters the combinations

i) $\chi_{i,j}^{\prime} = \sum_{\sigma} \langle f_{i \sigma}^{1
\dag} f_{j \sigma}^{2} \rangle = \sum_{\sigma} \langle f_{i
\sigma}^{2 \dag} f_{j \sigma}^{1} \rangle$: interplane uniform
RVB of spinons

\vskip2mm

ii) $\Delta_{i,j}^{\prime} = \mbox{$\frac{1}{\sqrt{2}}$} \langle
f_{i \uparrow}^{1} f_{j \downarrow}^{2} - f_{i \downarrow}^{1}
f_{j \uparrow}^{2} \rangle = \mbox{$\frac{1}{\sqrt{2}}$} \langle
f_{i \uparrow}^{2} f_{j \downarrow}^{1} - f_{i \downarrow}^{2}
f_{j \uparrow}^{1} \rangle$: interplane singlet

iii) $B_{i,j}^{\prime} = \langle b_{i \sigma}^{1} b_{j \sigma}^{2
\dag} \rangle = \langle b_{i \sigma}^{2} b_{j \sigma}^{1 \dag}
\rangle$: interplane uniform RVB of holons.

\smallskip
\noindent
For historical reasons \cite{ram}, the terminology ``uniform RVB''
is taken to signify the hopping order parameter for spin or charge,
whose finite value signifies diagonal coherence of the motion.
As will be shown below, for the spin order parameters it is sufficient
to consider only those between nearest-neighbour sites, so that
$\tau = \pm {\bf x}$ or $\pm {\bf y}$ for $\chi$ and $\Delta$, and
$\tau = \pm r_c {\bf z}$ for $\chi^{\prime}$ and $\Delta^{\prime}$.
In the single-layer problem, the free
energy is minimised when $\chi_{i,i+\tau}$ is uniform, but when
$\Delta_{i,i \pm x} = - \Delta_{i,i \pm y}$ \cite{rgjr,rkl}, so that
in reciprocal space the gap function $\Delta_k$ has $d$-wave symmetry.

	With these definitions, the mean-field approximation to the
Hamiltonian (\ref{eh}) in reciprocal space can be represented as
\begin{equation}
H = H_0 + \sum_k {\bf b}_{k}^{\dag} {\bf {\, M}}_{b k} {\bf \, b}_k
+ \sum_k {\bf f}_{k}^{\dag} {\bf {\, M}}_{f k} {\bf {\, f}}_k ,
\label{emh}
\end{equation}
where $H_0$ contains terms quadratic in the order parameters, and
${\bf b}_{k}^{\dag} = \left( b_{k}^{1 \dag}, b_{k}^{2 \dag} \right)$
and ${\bf f}_{k}^{\dag} = \left( f_{k \uparrow}^{1 \dag}, f_{-k
\downarrow}^{1}, f_{k \uparrow}^{2 \dag}, f_{-k \downarrow}^{2}
\right)$ are the boson and fermion state vectors. The boson
and fermion matrices are given by
\begin{equation}
{\bf M}_{b k} = \left( \begin{array}{cc}
- \omega_k & - \mbox{$\frac{1}{2}$} t_{\perp}^0 \chi_{k}^{\prime
\ast} \\ - \mbox{$\frac{1}{2}$} t_{\perp}^0 \chi_{k}^{\prime} &
- \omega_k \end{array} \right),
\label{ebhm}
\end{equation}
and
\begin{equation}
{\bf M}_{f k} = \left( \begin{array}{cccc}
- \xi_k & \Delta_k & - M_{-k} & D_{-k} \\
\Delta_k & \xi_k & D_{k}^{\ast} & M_{k}^{\ast}  \\
- M_{-k}^{\ast} & D_{k} & - \xi_k & \Delta_{k} \\
D_{-k}^{\ast} & M_k & \Delta_{k} & \xi_k
\end{array} \right)
\label{efhm}
\end{equation}
with the following definitions. The band energies for the
quasiparticles are $\xi_k = - [t_{fk} + t_{fk}^{\prime}
+ t_{fk}^{\prime\prime} + \mbox{$\frac{3}{4}$} J \chi_k]
- \mu$ and $\omega_k = - [t_{bk} + t_{bk}^{\prime} + t_{bk}
^{\prime\prime}] - \lambda$, in which the in-plane hopping terms are
$t_{fk}^{\alpha} = \sum_{k^\prime} t_{k^\prime - k}^{\alpha} \langle
b_{k^\prime}^{\dag} b_{k^\prime} \rangle$ and $t_{bk}^{\alpha} =
\sum_{k^\prime \sigma} t_{k^\prime - k}^{\alpha} \langle
f_{k^\prime \sigma}^{\dag} f_{k^\prime \sigma} \rangle$, with
$t_{k}^{\alpha} = \sum_{\tau_{\alpha}} t^{\alpha} e^{i {\bf
k.\tau}}_{\alpha}$, and $\mu$ and $\lambda$ are fermion and boson
chemical potentials. The $d$-symmetric gap is given by $\Delta_k
= \mbox{$\frac{3 \sqrt{2}}{4}$} J \Delta ( \cos k_x - \cos k_y )$.
The effective uniform interplane order parameter is $M_k =
t_{\perp k} B_{k}^{\prime} + \mbox{$\frac{3}{8}$} J_{\perp}
\chi_{k}^{\prime}$, where the $k$ dependence of $t_{\perp k}$ will
be discussed in detail below, and the interplane gap parameter is
$D_k = \mbox{$\frac{3 \sqrt{2}}{8}$} J_{\perp} \Delta_{k}^{\prime}$.

	The 4$\times$4 unitary matrix which diagonalises the
Hamiltonian of the fermionic part may be written as
\begin{equation}
{\bf T}_{k}^{-1} = \frac{1}{\sqrt{2}} \left( \begin{array}{cccc}
\cos \theta_{k}^{+} & \sin \theta_{k}^{+} & - e^{- i \phi} \cos
\theta_{k}^{-} & - e^{- i \phi} \sin \theta_{k}^{-} \\
- \sin \theta_{k}^{+} & \cosh \theta_{k}^{+} & e^{- i \phi} \sin
\theta_{k}^{-} & - e^{- i \phi} \cos \theta_{k}^{-} \\
e^{i \phi} \cos \theta_{k}^{+} & e^{i \phi} \sin \theta_{k}^{+}
& \cos \theta_{k}^{-} & \sin \theta_{k}^{-} \\
- e^{i \phi} \sin \theta_{k}^{+} & e^{i \phi} \cos \theta_{k}^{+}
& - \sin \theta_{k}^{-} & \cosh \theta_{k}^{-}
\end{array} \right) ,
\label{edm}
\end{equation}
where
\begin{equation}
\cos 2 \theta_{k}^{\pm} = \frac{\xi_k \pm \left| M_k \right|}
{E_{k}^{\pm}} \;\;\; {\rm {and}} \;\;\; \sin 2 \theta_{k}^{\pm} =
\frac{ \Delta_k \pm \left| D_k \right|}{E_{k}^{\pm}}.
\label{escp}
\end{equation}
The phase $\phi = - q_z r_c$ is chosen to cancel the $q_z$-dependent
phase of the interplane order parameters, and will be the same for
both $M_k$ and $D_k$; the modulus in $|Q_k|$ thus denotes the absence
of this phase, but does not prevent the order parameter from being
negative. It is the transformation (\ref{edm}) which will be used
extensively in the sections to follow, when the boson operators
${\bf f}_k$ are replaced in the physical quantities to be computed
by the quasiparticle operators ${\bf {\vec{\gamma}}}_k = {\bf T}_k
{\bf f}_k$.

	The fermion part of the Hamiltonian then takes the diagonal
form
\begin{equation}
\widetilde{\bf M}_{f k} = \left( \begin{array}{cccc}
- E_{k}^{+} & 0 & 0 & 0 \\ 0 & E_{k}^{+} & 0 & 0 \\
0 & 0 & - E_{k}^{-}  & 0 \\ 0 & 0 & 0 & E_{k}^{-}
\end{array} \right),
\label{edbhm}
\end{equation}
in which the eigenvalues are given by the combinations
\begin{equation}
E_{k}^{\pm} = \left[ \left( \xi_k \pm \left| M_k \right|
\right)^{2} + \left( \Delta_k \pm \left| D_k \right| \right)^{2}
\right]^{1/2} .
\label{eme}
\end{equation}
$E_{k}^{-}$ corresponds to the bonding band of the bilayer system,
and $E_{k}^{+}$ to the antibonding band.

	The 2 $\times$ 2 bosonic matrix (\ref{ebhm}) is trivially
diagonalised by the transformation
\begin{eqnarray}
b_{k}^{1} &=& \mbox{$\frac{1}{\sqrt{2}}$} \left( \beta_{k}^{1} -
e^{- i \phi} \beta_{k}^{2} \right) \label{esfcf} \\ b_{k}^{2} &=&
\mbox{$\frac{1}{\sqrt{2}}$} \left( e^{i \phi}
\beta_{k}^{1} - \beta_{k}^{2} \right)
\end{eqnarray}
to new quasihole operators ${\bf \beta}_k$, in terms of which
the diagonal boson matrix is
\begin{equation}
\widetilde{\bf M}_{b k} = \left( \begin{array}{cc}
- \widetilde{\omega}_{k}^{+} & 0 \\ 0 & - \widetilde{\omega}_{k}^{-}
\end{array} \right),
\label{edfhm}
\end{equation}
where the eigenenergies are given by
\begin{equation}
\widetilde{\omega}_{k}^{\pm} = \omega_k \pm \mbox{$\frac{1}{2}$}
t_{\perp}^0 \left| \chi_{k}^{\prime} \right|.
\label{efd}
\end{equation}

	The system is required to satisfy eight mean-field equations,
three for the in-plane order parameters, three for the interplane
parameters and two for the carrier number $\delta$. These equations
may be obtained by
minimising the free energy with respect to each of the parameters in
turn, or by substituting the quasiparticle operators into the
reciprocal-space definitions of the order parameters and transforming
back to real space. One obtains for the in-plane parameters
\begin{equation}
\chi_{i,i+\tau} = - \mbox{$\frac{1}{N}$} \sum_{k} e^{-i{\bf
{k.\tau}}} \left[ \cos 2 \theta_{k}^{+} \tanh \mbox{$\frac{1}{2}$}
\beta E_{k}^{+} + \cos 2 \theta_{k}^{-} \tanh \mbox{$\frac{1}{2}$}
\beta E_{k}^{-} \right] ,
\label{emfec}
\end{equation}
\begin{equation}
\Delta_{i,i+\tau} = - \mbox{$\frac{1}{\sqrt{2} N}$} \sum_{k} e^{-i{\bf
{k.\tau}}} \left[ \sin 2 \theta_{k}^{+} \tanh \mbox{$\frac{1}{2}$}
\beta E_{k}^{+} + \sin 2 \theta_{k}^{-} \tanh \mbox{$\frac{1}{2}$}
\beta E_{k}^{-} \right]
\label{emfeg}
\end{equation}
and
\begin{equation}
B_{i,i+\tau} = \mbox{$\frac{1}{N}$} \sum_{k} e^{-i{\bf {k.\tau}}}
\left[ n (\widetilde{\omega}_{k}^{+}) + n (\widetilde{\omega}_{k}
^{-}) \right] ,
\label{emfeb}
\end{equation}
for the interplane parameters
\begin{equation}
\chi_{i,i+\tau}^{\prime} = - \mbox{$\frac{1}{N}$} \sum_{k}
\left[ \cos 2 \theta_{k}^{+} \tanh \mbox{$\frac{1}{2}$}
\beta E_{k}^{+} - \cos 2 \theta_{k}^{-} \tanh \mbox{$\frac{1}{2}$}
\beta E_{k}^{-} \right] ,
\label{emfecp}
\end{equation}
\begin{equation}
\Delta_{i,i+\tau}^{\prime} = - \mbox{$\frac{1}{\sqrt{2} N}$} \sum_{k}
\left[ \sin 2 \theta_{k}^{+} \tanh \mbox{$\frac{1}{2}$}
\beta E_{k}^{+} - \sin 2 \theta_{k}^{-} \tanh \mbox{$\frac{1}{2}$}
\beta E_{k}^{-} \right]
\label{emfegp}
\end{equation}
and
\begin{equation}
B_{i,i+\tau}^{\prime} = \mbox{$\frac{1}{N}$} \sum_{k} e^{-i{\bf
{k.\tau}}} \left[ n (\widetilde{\omega}_{k}^{+}) - n
(\widetilde{\omega}_{k}^{-}) \right] ,
\label{emfebp}
\end{equation}
and for the carrier number, which must be equal to the number of
bosons, and by conservation must add to the number of fermions to
give unity,
\begin{equation}
\delta = \mbox{$\frac{1}{2 N}$} \sum_{k} \left[ n
(\widetilde{\omega}_{k}^{+}) + n (\widetilde{\omega}_{k}^{-}) \right]
\label{emfedb}
\end{equation}
from the boson part and
\begin{equation}
\delta = \mbox{$\frac{1}{2 N}$} \sum_{k}
\left[ \cos 2 \theta_{k}^{+} \tanh \mbox{$\frac{1}{2}$}
\beta E_{k}^{+} + \cos 2 \theta_{k}^{-} \tanh \mbox{$\frac{1}{2}$}
\beta E_{k}^{-} \right]
\label{emfedf}
\end{equation}
from the fermion part. In (\ref{emfec} - \ref{emfedf}), the sine and
cosine factors are given by (\ref{escp}) and $n (\omega_k)$ is
the Bose occupation function.

	In this type of model, the bosonic holons occupy only a very
small region of reciprocal space close to ${\bf k} = 0$, and in all
respects are effectively condensed at all reasonable temperatures.
Thus to a very good approximation one may take $B = \delta$ in
(\ref{emfeb}) and $B^{\prime} = \delta {\rm {sgn}}(\chi^{\prime})$ in
(\ref{emfebp}), so that the bosonic equations need not be considered
further. It is this approximation which allows us to consider only
nearest-neighbour spin order parameters in a model with only
nearest-neighbour superexchange interactions.
Thus the problem reduces to a five-parameter one in the
fermionic degrees of freedom, with the mean-field equations
(\ref{emfec}), (\ref{emfeg}), (\ref{emfecp}), (\ref{emfegp}) and
(\ref{emfedf}).

	Following Tanamoto {\it {et al.}} \cite{rTKFs,rTKF}, we
concentrate on a doping level close to the value where the system
is close to reproducing the properties of optimally-doped materials,
and choose $\delta = 0.2$. Here we expect a solution with finite
intra- and interplane uniform order parameters over a considerable
range of temperatures, and this is borne out by calculations. In the
low-temperature regime we wish to investigate the nature of the
transition to a state with singlet order, and the possible
coexistence or competition of the intra- and interplane ordered
states which may exist here. The idea of a singlet state with coupled
order parameters has been studied in one version of the current model
by Ubbens and Lee \cite{rul}, and proposed by Kuboki and Lee
\cite{rkkl} as a candidate which may explain recent observations of
the anisotropic gap in BSCCO materials by photoemission \cite{rag}.
A very recent numerical study \cite{reom} has investigated the same
issue in small, bilayer clusters, finding a sharp crossover from
predominantly in-plane to predominantly interplane singlet order at a
value of the interlayer superexchange parameter $J_{\perp} = x J$,
where $x$ is of order unity, but may under certain circumstances be
quite small. We will compare our results with those of Ref.
\cite{reom} in more detail below.

	In a single-layer system, the gap parameter $\Delta$ may have
an arbitrary phase $\theta$, from which no physical consequences arise.
In a bilayer, the phases of the parameters on each layer may differ,
thus appearing as $\Delta_k e^{i \theta_1}$ and $\Delta_k e^{i
\theta_2}$. When the layers are coupled, there may in general be an
additional term in the Hamiltonian of the form $H = - \epsilon \cos
(\theta_1 - \theta_2)$, where the energy scale of the phase coupling
parameter is $\epsilon \sim t_{\perp}^{0 \; 2} / W$, in which $W$ is
the bandwidth of the in-plane dispersion. This term may couple to an
external magnetic field, and gives rise to a variety of Josephson
coupling phenomena, some of which are integral to the qualitatively
new predictions of the interlayer pair tunnelling theory \cite{raipt}.
In this work, we assume that the minimum energy state is maintained,
so that the phases in the two layers remain equal, and both gap
functions can be taken simply to be $\Delta_k$, as given below
(\ref{efhm}).

	In order to make the following study more directly
applicable to the physical systems, we appeal to the results of
Andersen {\it {et al.}} \cite{roka} for some additional detail
about the interplane hopping term. From detailed bandstructure
studies on the YBCO system, these authors estimate the magnitude
to be $t_{\perp}^0 \simeq 0.15 t$, where in our choice of units
the maximal band splitting will be $8 t_{\perp}^0$. The $k$
dependence resulting from the combination of the interplane term
with the extended in-plane terms is contained consistently in the
extended interplane hopping parameters $t_{\perp}$,
$t_{\perp}^{\prime}$ and $t_{\perp}^{\prime\prime}$ (Fig.~1),
and is such that $t_{\perp} ({\bf k})$ is small along the
$\Gamma$M line in the Brillouin zone. This has been parameterised
in the form
\begin{equation}
t_{\perp} ({\bf k}) = t_{\perp}^0 \left( \cos k_x - \cos k_y
\right)^2,
\label{etpk}
\end{equation}
which corresponds to the special case where the bands are degenerate
along $\Gamma$M, and results from a real-space distribution of the
interplane hopping integrals given by $t_{\perp} = 0$ \cite{roka},
$t_{\perp}^{\prime} = - \mbox{$\frac{1}{2}$} t_{\perp}^0$ and
$t_{\perp}^{\prime\prime} = \mbox{$\frac{1}{4}$} t_{\perp}^0$.
The interlayer spin exchange is taken to be nearest-neighbour only,
and to have a small value $J_{\perp} \sim 0.1 J$, in line with the
majority of experimental \cite{rjpv1,rjpv2,rjpvs} and theoretical
\cite{roka} indications. We note, however, that recent infrared
transmission and reflectometry measurements \cite{rkrg} suggest a
considerably larger value of this parameter in the insulating YBCO
system, and so we do not discount this possibility.

	To investigate the nature of the spin-ordered state, we
consider the linearised gap equations (\ref{emfeg}) and
(\ref{emfegp}) close to the their transition temperature. In their
most general form these contain the 3 equations
\begin{eqnarray}
\Delta_x & = & \frac{3}{4} \sum_k \cos k_x \left( \frac{ \Delta_x
\cos k_x + \Delta_y \cos k_y + \mbox{$\frac{1}{2}$} J_{\perp}
\Delta_{\perp}}{2 \xi_{k}^{+}} \tanh \mbox{$\frac{1}{2}$} \beta
\xi_{k}^{+} \right. \label{elg} \nonumber \\ & & \;\;\;\;\;\;\;\;
\;\;\;\;\;\;\;\; + \left. \frac{ \Delta_x \cos k_x + \Delta_y \cos
k_y - \mbox{$\frac{1}{2}$} J_{\perp} \Delta_{\perp}}{2 \xi_{k}^{-}}
\tanh \mbox{$\frac{1}{2}$} \beta \xi_{k}^{-} \right) \nonumber \\
\Delta_y & = & \frac{3}{4} \sum_k \cos k_y \left( \frac{ \Delta_x
\cos k_x + \Delta_y \cos k_y + \mbox{$\frac{1}{2}$} J_{\perp}
\Delta_{\perp}}{2 \xi_{k}^{+}} \tanh \mbox{$\frac{1}{2}$} \beta
\xi_{k}^{+} \right. \nonumber \\ & & \;\;\;\;\;\;\;\;\;\;\;\;
\;\;\;\; + \left. \frac{ \Delta_x \cos k_x + \Delta_y \cos k_y -
\mbox{$\frac{1}{2}$} J_{\perp} \Delta_{\perp}}{2 \xi_{k}^{-}}
\tanh \mbox{$\frac{1}{2}$} \beta \xi_{k}^{-} \right) \\
\Delta_{\perp} & = & \frac{3}{4} \sum_k \left( \frac{ \Delta_x
\cos k_x + \Delta_y \cos k_y + \mbox{$\frac{1}{2}$} J_{\perp}
\Delta_{\perp}}{2 \xi_{k}^{+}} \tanh \mbox{$\frac{1}{2}$} \beta
\xi_{k}^{+} \right. \nonumber \\ & & \;\;\;\;\;\;\;\;\;\;\;\;
\;\;\;\; - \left. \frac{ \Delta_x \cos k_x + \Delta_y \cos k_y -
\mbox{$\frac{1}{2}$} J_{\perp} \Delta_{\perp}}{2 \xi_{k}^{-}}
\tanh \mbox{$\frac{1}{2}$} \beta \xi_{k}^{-} \right) , \nonumber
\end{eqnarray}
where $\xi_{k}^{\pm} = \xi_k \pm |M_k|$ and $\Delta_{\perp}$
denotes $\Delta^{\prime}$. This set of equations can be cast in
the schematic matrix form
\begin{equation}
\left( \begin{array}{ccc} I_{1}^{xx} + I_{2}^{xx} - 1 & I_{1}^{xy}
+ I_{2}^{xy} & I_{1}^{x} - I_{2}^{x}  \\  I_{1}^{xy} + I_{2}^{xy}
& I_{1}^{yy} + I_{2}^{yy} - 1 & I_{1}^{y} - I_{2}^{y} \\ I_{1}^{x}
- I_{2}^{x} & I_{1}^{y} - I_{2}^{y} & I_1 + I_2
- \mbox{$\frac{2}{J_{\perp}}$} \end{array} \right) \left(
\begin{array}{c} \Delta_x \\ \Delta_y \\ \mbox{$\frac{1}{2}$}
J_{\perp} \Delta_{\perp} \end{array} \right) = \left(
\begin{array}{c} 0 \\ 0 \\ 0 \end{array} \right) ,
\label{emlg}
\end{equation}
where $I_{\alpha} = \langle 1 \rangle_{\alpha}$,
$I_{\alpha}^{i} = \langle \cos k_i \rangle_{\alpha}$ and
$I_{\alpha}^{ij} = \langle \cos k_i \cos k_j \rangle_{\alpha}$, in
which the angled brackets $\langle$ $\rangle_{\alpha}$ denote
integration over the Brillouin zone of the energy denominator and
thermal function of the energy branch $\alpha$. A coupled
transition of all 3 order parameters may occur where the determinant
of this matrix is zero. The determinantal equation may be
represented simply by
\begin{equation}
\left| \begin{array}{ccc} a & b & c \\ b & a & c \\ c & c & d
\end{array} \right| = (a - b) \left[ ( a + b ) d - 2 c^2 \right] = 0 .
\label{edete}
\end{equation}

	The in-plane $d$-symmetric solution is sought by writing the
eigenvector $\left( \Delta_x, \Delta_y, \Delta_{\perp} \right)$ as
$\left( \Delta, - \Delta, \Delta_{\perp} \right)$, where $\Delta$ is
real, and it is clear that the only possible solution is
$\Delta_{\perp} = 0$. This is the root $a - b = 0$ of (\ref{edete})
above. Thus in the mean-field framework, $d$-wave singlet pairing in
the plane is thus found to exclude any interplane singlet order at
the same onset temperature $T = T_{RVB}$. This is the most
important qualitative result of the current analysis, as it has
profound implications for the nature of the singlet spin-ordered
state which may be realised in this type of model. The in-plane
$s$-wave solution $\left( \Delta, \Delta, \Delta_{\perp} \right)$
can be found to exist with $\mbox{$\frac{\Delta_{\perp}}{\Delta}$}
= - \mbox{$\frac{2 c}{d}$}$, which is the root $[ \;\;\;\; ] = 0$
in (\ref{edete}) above. This is a coupled solution with in-plane
$s$-wave symmetry, which for small $J_{\perp}$ admits some admixed
interplane singlet order in an amount proportional to the band
splitting, and at large $J_{\perp}$ represents a predominantly
interplane singlet-ordered state. We have not been
able to find a solution to the system of mean-field equations for
the in-plane $s$-wave ordered state for any values of the doping
$\delta$ or interplane superexchange interaction $J_{\perp}$
in the regime of physical interest, and so do not consider this state
further here. We will have some further comment on the interplane
ordered state below, in the context of its possible coexistence with
in-plane $d$-wave singlet pairing at temperatures away from $T_{RVB}$.

	Returning to consider the case of a $d$-symmetric in-plane
singlet order parameter in more detail, we note first that if the
planes are decoupled ($t_{\perp}^0 = 0$), $\xi_{k}^{+} = \xi_{k}^{-}$
and the equations for $(\Delta_x, \Delta_y)$ and $\Delta_{\perp}$
decouple. Thus there would be two separate transitions for the total
order parameter if interplane singlet ordering were not excluded as
in the coupled case above. To investigate the possibility of
coexisting order parameters in some temperature regime of the system,
one may consider the in-plane $d$-symmetric ordered state and seek a
second transition to finite interplane order at a lower temperature.

	In the absence of interplane singlet order, this state is
described by the reduced system of 4 equations ((\ref{emfec}),
(\ref{emfeg}), (\ref{emfecp}) and (\ref{emfedf})), which can be solved
readily to yield a self-consistent set of solutions for the
order parameters as functions of temperature. In Fig.~2 we show the
transition temperature $T_{RVB}$, representing the onset of in-plane
singlet order in such a system, as a function of the interplane
hopping parameter $t_{\perp}^0$. $T_{RVB}$ is little affected by small
interplane hopping, but loses most of its value as $t_{\perp}^0$
(\ref{etpk}) approaches $J$. These results were obtained with a
fixed, small interplane spin interaction $J_{\perp} = 0.1 J$ at all
values of $t_{\perp}^0$.

	Returning to the system of 5 equations, this can be studied
below $T_{RVB}$ at several values of $t_{\perp}^0$ (Fig.~2) to seek a
transition to non-zero $\Delta_{\perp}$ (\ref{emfegp}). We find that
this possibility exists only for values of $0.8 J < J_{\perp} < 1.2 J$:
for smaller values the equation cannot be satisfied because the
coefficient $J_{\perp}$ of the left-hand side is too small to compete
with the band splitting, and for larger ones it is likely that the
assumption of the interplane singlet state being subsidiary to the
in-plane state is no longer valid \cite{reom}. The range of
coexistence is quite insensitive to the value of $t_{\perp}^0$, a
feature which we believe is due to the line nodes in $t_{\perp k}$
(\ref{etpk}). This result is in agreement with a recent study by Eder
{\it {et al.}} \cite{reom}, who investigated spin correlation
functions in 16- and 20-site bilayer clusters by exact
diagonalisation. These authors find for a doped system a very abrupt
crossover, from a state with predominantly in-plane $d$-wave spin
correlation to one of ``standing singlets'' formed between
nearest-neighbour spins in each layer, as a function of increasing
interplane superexchange $J_{\perp}$. The crossover occurs at a value
$J_{\perp} = x J$, in which $x$ varies over a parameter-dependent
range of values below unity. That interplane singlet order is not
compatible with in-plane $d$-symmetric singlet pairing was also found
by Liechtenstein {\it {et al.}} \cite{rlma} in a bilayer
antiferromagnetic spin fluctuation model, which was shown to favour
the existence of the anisotropic $s$-wave state.

	These results raise the interesting question of how the
$d$-symmetric in-plane singlet order in an isolated plane may evolve
as interplane hopping and superexchange increase. One alternative to
the admixture of real order parameters considered above, which has
been shown not to evolve below some significant threshold value of
the coupling, is the development of imaginary components of the order
parameter. If the order parameter has the form $\Delta_x = \Delta_1
+ i \Delta_2$, $\Delta_y = - \Delta_1 + i \Delta_2$, one may
formulate a system of equations analogous to (\ref{elg}) above, in
which the real part is the single-layer equation for in-plane
$d$-order and the imaginary part a coupled equation for $\Delta_2$
and a purely imaginary interplane component $i \Delta_{\perp}$.
This has been termed the $d + is$ state, and would give a completely
nodeless gap because the order parameter has a finite imaginary part
where the real part vanishes. However, the real and imaginary parts
are decoupled, so there is no reason to expect the two transitions to
occur simultaneously. A second transition below $T_{RVB}$ to a nodeless
state has been proposed as the explanation for a feature in the NMR
spin-echo relaxation rate $T_{2G}^{-1}$ \cite{riphd}, and so this
state is of both experimental and theoretical interest. We have
sought a transition to this state in the same way as described above,
and again found that it did not exist within the present framework
below values of $J_{\perp} \sim J$.

	A different proposition for the existence of the coupled
singlet state is given by the argument \cite{rul,rilma} that
fluctuations of the gauge-field in a bilayer system are considerably
more effective in suppressing in-plane $d$-wave order than interplane
$s$-wave order, and thus act to allow the latter to survive in the
presence of the former. The same authors also argue for enhancement
of the interlayer coupling $J_{\perp} ({\bf q})$, to values which
may fall in the coupled regime, close to the antiferromagnetic
wavevector $(\pi,\pi)$. Finally,
if the system is orthorhombic, {\it {i.e.}} $t_x \ne t_y$ and
$J_x \ne J_y$ in the plane, one may show \cite{rkkl,rkpc} that the
possibility of a coupled transition does exist, and thus that a
mixed ``$d + s$'' or ``$d + is$'' state of coexisting intra- and
interplane order parameters is justifiable in the physical parameter
regime. However, there is as yet no definitive evidence that the
BSCCO system, whose photoemission spectra have been interpreted to
suggest a $d + s$ state \cite{rkkl}, shows any orthorhombicity of a
type which would couple appropriately to both gap parameters.

	The range of $J_{\perp}$ for which a mixed state of in- and
interplane singlet spin order may exist is unrealistically large
for the high-$T_c$ superconducting materials, and in addition is
unphysical in the context of a model where the superexchange
interaction should be the result of a second-order hopping process,
with magnitude $J_{\perp} \sim \frac{t_{\perp}^{0 \; 2}}{U}$. In
order to investigate the properties to be
expected from the model, and to compare them with experiment, we wish
to restrict the discussion to the physical parameter regime, which
will require considerably smaller values of $J_{\perp}$. We will
thus proceed by taking the interplane singlet order parameter to be
zero, and to study the 4-parameter problem for the uniform in- and
interplane order parameters $\chi$ and $\chi^{\prime}$, the in-plane
spin gap parameter $\Delta$ and the chemical potential $\mu$, albeit
in the presence of a finite interplane spin coupling constant.

	Following Andersen {\it {et al.}}
\cite{roka}, we now fix the magnitude of
the interplane hopping term at $t_{\perp}^0 = 0.15 t$. We choose to
set the interplane spin coupling to the small value $J_{\perp} =
0.1 J$, noting that if in fact this parameter is found by more
detailed experiments to have a somewhat larger value, its role in
the structure of the current theory is such that it has a
negligible effect on the mean-field solutions and Fermi surface
shape. In the spirit of the single-layer analysis of Tanamoto and
coworkers \cite{rTKFs,rTKF}, we wish to choose the extended transfer
integrals $t^{\prime}$ and $t^{\prime\prime}$ in such a way that
the Fermi surfaces of the bonding and antibonding bands reproduce
as many features as possible of the physical bilayer systems of
interest, BSCCO and YBCO. As was shown clearly by these authors,
the magnetic properties of the system are strongly influenced by
the exact shape of the Fermi surface.

	As a prototype of the BSCCO system, in Fig.~3(a) we show
Fermi surfaces in the first quadrant of reciprocal space for the
parameter choice $t = 4 J$, $t^{\prime} = - \mbox{$\frac{1}{8}$}
t$ and $t^{\prime\prime} = \mbox{$\frac{1}{6}$} t$. For the fixed
value of $t_{\perp}^0$,
this leads to a pair of dispersion surfaces with flat regions
around the $(\pi,0)$ saddle points, the important feature of which
is that the ``extended saddle point'' regions of the bonding band
lie below the chemical potential, while those of the antibonding
band lie above it. Thus the bonding (outer) band has the open
shape characteristic of YBCO-like systems \cite{rTKF}, while the
antibonding (inner) band has the closed, LSCO-like shape; we will
use the terms ``open'' and ``closed'' to refer to the nature of the
Fermi surface in the first Brillouin zone, in an electron picture.
In addition, the parameters
are chosen so that the antibonding band is rather strongly nested,
by which is meant that the Fermi surface has significant regions
lying almost on a straight line parallel to $q_x + q_y = 0$, so
that the same wavevector ${\bf q}$ can span many pairs of Fermi
surface points; in detail, this surface is slightly concave in
relation to the given line. Finally, the bands appear to be
degenerate along the direction $\Gamma$M in the Brillouin zone, in
accordance with the calculations of Ref. \cite{roka}, and so we
retain the form for $t_{\perp k}$ (\ref{etpk}) proposed by these
authors. These are the features of the published interpretation of
the photoemission data from BSCCO in Ref. \cite{rsg}, and so we take
this parameter set to characterise the BSCCO system in the remainder
of this study.

	To describe the YBCO system, in Fig.~3(b) we show Fermi
surfaces in the first quadrant for the parameter choice $t = 4 J$,
$t^{\prime} = - \mbox{$\frac{1}{5}$} t$ and $t^{\prime\prime} =
\mbox{$\frac{1}{4}$} t$. In this case the parameters are such
that the extended saddle point regions of both bands are below the
chemical potential, so that both have open Fermi surfaces. However,
the antibonding band is very flat in these regions, and the Fermi
surface shows a rather narrow ``neck'', so is close to becoming
closed on small changes of the parameters. We emphasise in passing
that these strong alterations in Fermi surface shape are a
consequence of very small changes in the parameters $t^{\prime}$
and $t^{\prime\prime}$, because of the flat nature of the
saddle-point region. Here we require that the bands are not fully
degenerate in the $\Gamma$M direction, and so alter the form of
the interplane term to $t_{\perp k} = \mbox{$\frac{2}{3}$} \left[
\left( \cos k_x - \cos k_y \right)^2 + 2 \right]$. This alteration
corresponds to the case where the interplane hopping integrals
are given by $t_{\perp} = 0$, $t_{\perp}^{\prime} =
- \mbox{$\frac{1}{3}$} t_{\perp}^0$ and $t_{\perp}^{\prime\prime}
= \mbox{$\frac{1}{6}$} t_{\perp}^0$, or in other words that the
interplane hopping is less extended in this case. The function in
reciprocal space remains minimal along the $\Gamma$M.
The fitted features are those of the YBCO
system, as calculated from first principles in Ref. \cite{roka},
and are fully consistent with results from the most accurate
photoemission studies of YBCO 124 \cite{ra124} and 123
\cite{ra123} to date. Henceforth we take this parameter set to
give the prototypical YBCO system for analysis of the physical
properties. We believe that the relatively large value of
$t^{\prime\prime} = \mbox{$\frac{1}{4}$} t$ is reasonable, as it
remains consistent with the expectation from microscopic models for
the transfer of extended quasiparticles in the CuO$_2$ plane
\cite{rmfjef}. The Fermi surfaces shown in Fig.~3 are independent of
the perpendicular wavevector $k_z$, which does not appear in the
quasiparticle dispersion (\ref{eme}) due to the assumption of no
coupling between unit cells. In the
slave-boson decomposition at the mean-field level, because the
holons are concentrated at $q = 0$, calculated photoemission
spectra are exactly characteristic of the spinon Fermi surfaces
(Fig.~3), and contain no alterations reflecting any effects of the
charge degrees of freedom on the spin.

	With the above choices one may solve the mean-field
equations of the 4-parameter fermion system, to obtain all of the
parameters as functions of temperature. In Fig.~4 we show the form
of the in-plane singlet spin order parameter $\Delta(T)$ for both
BSCCO-like (Fig.~4(a)) and YBCO-like (Fig.~4(b)) systems. $\Delta$
shows a BCS-like second-order
transition, and the transition temperature $T_{RVB}$ sets the
characteristic scale for spin-related properties in this type of
model. We turn now to the computation of two such properties,
the dynamic susceptibility and the phonon anomalies.

\section{Dynamic Susceptibility}

	The dynamic susceptibility $\chi ({\bf q}, q_z, \omega)$
contains all information about the spin response of a magnetic
material, at all frequencies and wavevectors. Its zero-frequency
limit is sampled at certain wavevectors by NMR experiments, while the
full dynamical quantity is measured by inelastic neutron scattering,
as discussed in more detail from the viewpoint of this type of model
in Ref. \cite{rTKF}. Here we will compute the dynamic susceptibility
in the random-phase approximation (RPA), one method by which spin
fluctuation enhancement may be taken into account, and illustrate
the results as scans of in-plane wavevector ${\bf q}$ or frequency
$\omega$ with the other variable held fixed. We will show also the
variation of the susceptibility with the wavevector component $q_z$,
and explain the results in terms of intra- and interband scattering
processes.

	In a bilayer system, the imaginary part of the bare, retarded
susceptibility ${\bf {\chi}}_{0}^{\prime\prime} ({\bf q}, q_z,
\omega)$ is a $2\times2$ matrix which can be calculated by analytic
continuation of the associated thermal function
\begin{equation}
{\bf {\chi}}_{0} ({\bf q}, q_z, i \omega_n) =  \left( \begin{array}{cc}
\langle {\bf S}^1 {\bf .S}^1 \rangle & \langle {\bf S}^1 {\bf .S}^2
\rangle \\ \langle {\bf S}^2 {\bf .S}^1 \rangle & \langle {\bf S}^2
{\bf .S}^2 \rangle \end{array} \right),
\label{eimchit}
\end{equation}
in which $\langle {\bf S}^l {\bf .S}^m \rangle$ denotes $\langle {\bf
S}^l ({\bf q}, q_z, i \omega_n) {\bf .S}^m ({\bf q}, q_z, i \omega_n)
\rangle$. The spin operator ${\bf S}$ may be written in terms of the
operators $\vec{f}_k$, transformed to the quasiparticle operators
$\vec{\gamma}_k$ and the resulting expressions evaluated in the
diagonal basis. Omitting details of this lengthy procedure, the bare
susceptibility assumes the form
\begin{equation}
{\bf {\chi}}_{0} ({\bf q}, q_z, \omega) =  \left(
\begin{array}{cc}
\chi_{+} ({\bf q}, \omega) & e^{i q_z r_c} \chi_{-} ({\bf q},
\omega) \\ e^{- i q_z r_c} \chi_{-} ({\bf q}, \omega) &
\chi_{+} ({\bf q}, \omega) \end{array} \right),
\label{eimchir}
\end{equation}
where the diagonal and off-diagonal parts are quantities whose
imaginary parts have the form
\begin{eqnarray}
\chi_{\pm}^{\prime\prime} ({\bf q}, \omega) & = & \sum_{\alpha
\beta} \frac{\pi}{8} A^{\pm} \sum_k \left[ C_{k,q}^{+} T_{k,q}^{+}
\left[ \delta \left( \omega - E_{k}^{\beta} + E_{k+q}^{\alpha}
\right) - \delta \left( \omega + E_{k}^{\beta} - E_{k+q}^{\alpha}
\right) \right] \right. \label{echi0} \nonumber \\ & & \;\;\;\;\;\;
\;\;\;\;\;\;\;\;\;\;
+ \left. C_{k,q}^{-} T_{k,q}^{-} \left[ \delta \left(
\omega + E_{k}^{\beta} + E_{k+q}^{\alpha} \right) - \delta \left(
\omega - E_{k}^{\beta} - E_{k+q}^{\alpha} \right) \right] \right] ,
\end{eqnarray}
and similarly for the real parts $\chi_{\pm}^{\prime}$ \cite{rlw}.
Here the indices $\alpha$ and $\beta$
denote the bonding or antibonding bands, $A^{+} = 1$ and $A^{-} = 2
\delta_{\alpha \beta} - 1$,
\begin{equation}
C_{k,q}^{\pm} = 1 \pm \cos 2 \left(
\theta_{k+q}^{\alpha} - \theta_{k}^{\beta} \right)
\label{enacf}
\end{equation}
gives the coherence factors and
\begin{equation}
T_{k,q}^{\pm} = f( \pm E_{k}^{\beta}) - f(E_{k+q}^{\alpha})
\label{enatf}
\end{equation}
gives the thermal factors. Thus each component of ${\bf {\chi}}_{0}$
is a combination of 16 terms, 8 for normal scattering processes and
8 for anomalous ones. All of the contributions are additive in
$\chi_{+}$, but the interband contributions ($\alpha \ne \beta$)
are subtractive in $\chi_{-}$.

	Note that the only dependence on $q_z$ is that contained
explicitly in the phase factors of the off-diagonal terms in
(\ref{eimchir}). This phase will appear in the imaginary part as the
legitimate coefficient of terms proportional to a $\delta$-function
in energy which are responsible for dissipative processes, and is not
forbidden by the symmetry of ${\bf {\chi}}^{\prime\prime}$. On taking
the symmetric combinations required to compute any physical quantity,
the phases will combine to give a real result.

	The RPA susceptibility is a sum of chains of bare
polarisation parts connected by the spin interaction ${\bf J}
({\bf q}, q_z)$, and can be written in the form of a matrix Dyson
equation as
\begin{eqnarray}
{\bf {\chi}} (q, \omega) & = & {\bf {\chi}}_{0} (q, \omega) -
{\bf {\chi}}_{0} (q, \omega) {\bf J} (q) {\bf {\chi}} (q, \omega)
\label{echid} \\ & = & \left( {\bf 1} + {\bf {\chi}}_{0} (q, \omega)
{\bf J} (q) \right)^{-1} {\bf {\chi}}_{0} (q, \omega) ,
\label{echirpa}
\end{eqnarray}
where $q$ denotes $({\bf q}, q_z)$. The matrix form of the spin
interaction on the bilayer is
\begin{equation}
{\bf J} ({\bf q}, q_z) =  \left( \begin{array}{cc}
J_{\parallel} ({\bf q}) & e^{i q_z r_c} J_{\perp} ({\bf q}) \\
e^{- i q_z r_c} J_{\perp} ({\bf q}) & J_{\parallel} ({\bf q})
\end{array} \right),
\label{ejm}
\end{equation}
in which by the assumption of nearest-neighbour interactions only,
one has $J_{\parallel} ({\bf q}) = J (\cos q_x + \cos q_y)$ and
$J_{\perp} ({\bf q}) = J_{\perp}$.

	Omitting the functional dependences and superscripts for
clarity, the RPA susceptibility from (\ref{echirpa}) has the form
\begin{equation}
{\bf {\chi}} = \frac{1}{|det|} \left( \begin{array}{cc}
\chi_{+} + J_{\parallel} ( \chi_{+}^2 - \chi_{-}^2 ) & e^{i \phi}
\left( \chi_{-} + J_{\perp} ( \chi_{+}^2 - \chi_{-}^2 ) \right)
\\ e^{- i \phi} \left( \chi_{-} + J_{\perp} ( \chi_{+}^2 -
\chi_{-}^2 ) \right) & \chi_{+} + J_{\parallel} ( \chi_{+}^2 -
\chi_{-}^2 ) \end{array} \right) ,
\label{echiem}
\end{equation}
where $\phi$ denotes $q_z r_c$ and the determinant of the matrix
$\left( {\bf 1} + {\bf {\chi}}_{0} {\bf J} \right)$ has the separable
form
\begin{eqnarray}
|det| & = & 1 + 2 \left( \chi_{+} J_{\parallel} + \chi_{-} J_{\perp}
\right) + \left( \chi_{+}^2 - \chi_{-}^2 \right) \left( J_{\parallel}
^2 - J_{\perp}^2 \right) \label{edet} \nonumber \\ & = & \left( 1 +
(J_{\parallel} + J_{\perp}) (\chi_{+} + \chi_{-}) \right) \left( 1 +
(J_{\parallel} - J_{\perp}) (\chi_{+} - \chi_{-}) \right) .
\end{eqnarray}
The measured quantity in an experiment such as the determination of
the neutron scattering cross section is the sum of all components
of the imaginary part of the susceptibility matrix, {\it {i.e.}}
\begin{equation}
S^R (q, \omega) \propto \sum_{\mu \nu} \frac{ \chi_{\mu \nu}
^{\prime\prime} (q, \omega)} { 1 - e^{- \beta \omega}} ,
\label{enscs}
\end{equation}
where $\mu$ and $\nu$ denote the layer index.
Thus the diagonal terms in (\ref{echiem}) are the effective RPA
susceptibility for spins in a single plane of the bilayer, and the
off diagonal terms give the interplane contribution, which will be
proportional to $\cos q_z r_c$.

	By performing the summation and separation, the final form
of the cross section is
\begin{equation}
S^R (q, \omega) \propto {\sl {Im}} \, \left\{ \frac{ \left( \chi_{+} +
\chi_{-} \right) \cos^2 \mbox{$\frac{1}{2}$} q_z r_c }{ 1 +
(J_{\parallel} + J_{\perp}) (\chi_{+} + \chi_{-}) } + \frac{
\left( \chi_{+} - \chi_{-} \right) \sin^2 \mbox{$\frac{1}{2}$}
q_z r_c }{ 1 + (J_{\parallel} - J_{\perp}) (\chi_{+} - \chi_{-}) }
\right\} ,
\label{esnscs}
\end{equation}
a more general version of the result given by Ref. \cite{rul}.
Returning to the equation (\ref{echi0}), one observes that the
components may be cast in the form $\chi_{\pm} = \chi_{\alpha
\alpha} \pm \chi_{\alpha \beta} \pm \chi_{\beta \alpha} +
\chi_{\beta \beta}$, where each of the terms $\chi_{\alpha
\alpha^{\prime}}$ has contributions from 2 normal and 2 anomalous
scattering processes, and is either purely intra- ($\alpha =
\alpha^{\prime}$) or interband. The 2 terms in (\ref{esnscs}) are
then either purely intraband $\left( \chi_{+} + \chi_{-} = 2 (
\chi_{\alpha \alpha} + \chi_{\beta \beta} ) \right)$ or interband
$\left( \chi_{+} - \chi_{-} = 2 ( \chi_{\alpha \beta} +
\chi_{\beta \alpha}) \right)$ in nature. Thus the wavevector $q_z$
may be used to select which type of scattering process is measured,
so that the contributions of each may be compared.

	In Figs.~5-7 we give results for the wavevector and
frequency dependence of the imaginary part of the dynamic
susceptibility $\chi = \sum_{\mu \nu} \chi_{\mu \nu}$ for the BSCCO-
and YBCO-like systems, based on the Fermi surface parameters chosen
in section 2. Here we will show only results for the system at low
temperatures (specifically, $T = 0.1 T_{RVB}$), so that it is in the
singlet RVB state; the qualitative features of the evolution from the
high-temperature state are well illustrated in Ref. \cite{rTKF}.
In Fig.~5(a) is shown $\chi^{\prime\prime} ({\bf q}, \omega)$ for
the BSCCO system as ${\bf q}$ is scanned along the symmetry
directions of the Brillouin zone. The frequency is fixed
at $\omega = 0.3 J$, above the value of the twice the maximal gap
$\Delta_k$, and the out-of-plane wavevector at
$q_z = 0$ so that only scattering processes within the bonding
and antibonding bands may contribute. One observes a significant
incommensuration due to the strongly-nested shape of the antibonding
band, and a general increase in the scattered intensity for
values of ${\bf q}$ near ${\bf Q} = (\pi,\pi)$, the antiferromagnetic
wavevector which spans many points close to the Fermi surface of the
bonding band (Fig.~3). This is a combination of the Fermi-surface
features documented in Ref. \cite{rTKF} for both closed and open
types, which clearly vary little with energy transfer.
In Fig.~6(a) is shown the same quantity for $q_z = \pi$ (in
``bilayer units'', $r_c = 1$), so that only bonding to antibonding
band scattering processes are sampled. Here the scattered intensity
is considerably higher, indicating a greater enhancement from the
relevant denominator in (\ref{esnscs}), and is largely commensurate
as a result of poor nesting between the two bands. In Fig.~7(a) is
shown $\chi^{\prime\prime} ({\bf q}, \omega)$ at fixed wavevector
${\bf q} = {\bf Q}$ as the frequency is scanned up to values of order
$J$, for 3 choices of $q_z$ and at low temperature. The results for
interband scattering are very similar to those obtained in Ref.
\cite{rTKF} for the single layer with open Fermi surface,
in that there is a strong peak due
to states pushed out of a gap ($\Delta_{\alpha} ({\bf k}_{\alpha})
+ \Delta_{\beta} ({\bf k}_{\beta})$ here, with ${\bf k}_{\alpha} -
{\bf k}_{\beta} = (\pi,\pi)$), followed by a broad region of
spectral weight up to a shoulder at $\omega \sim J$ which
characterises the maximal energy separation of significant numbers
of points whose wavevector separation is $(\pi,\pi)$ for the energy
dispersions of the system. For intraband scattering, the shoulder
appears at a similar energy, but the peak feature is missing; this
can be understood due to the fact that there is no contribution from
processes involving the antibonding band, all of whose spanning
wavevectors are smaller than ${\bf Q}$, with the result that the
denominator in the first term of (\ref{esnscs}) is far from being
small, and so provides very little enhancement.

	Fig.~5(b) shows the variation of $\chi^{\prime\prime}$ with
${\bf q}$ in the Brillouin zone at fixed frequency, low temperature
and $q_z = 0$ for the YBCO system, where it is clear that the spin
excitations are commensurate, and strong over a broad range of
${\bf q}$ around $(\pi,\pi)$. This result appears from \cite{rTKF}
to be characteristic of the open Fermi surface shape; the only
qualitative effect of the higher energy transfer shown here in the
singlet RVB state is that one no longer observes the peaks which
arise due to scattering processes between the gap nodes.
For $q_z = \pi$ (Fig.~6(b)), the peak is again commensurate, as is to be
expected as the 2 bands involved both have the open shape, and once
again has higher intensity than that from intraband processes.
Fig.~7(b) shows that the excitation spectrum as a function of $\omega$
once more has a shape similar to that seen in the single-layer case,
with the exception of clear evidence of 2 separate contributions to
the peaks at each $q_z$. This is the result of having 2 pairs of
Fermi-surface points separated by $(\pi,\pi)$, so that the
characteristic gap combinations in each case may differ, and is most
pronounced in the $q_z = \pi$ case. The shoulder feature for the YBCO
bands occurs at $\omega \sim 1.2 J$.

	We may conclude by noting that at the RPA level the dynamic
susceptibility is very much dominated by Fermi-surface scattering
processes, which are strongly favoured over even those involving
nearby points, and in this sense the results show little qualitative
difference from those of the single layer. The calculated result that
interplane contributions are stronger than intraplane ones over most
of the energy range (Fig.~7) arises primarily from their smaller
enhancing denominator (\ref{esnscs}), and is consistent with the
results of experiment in YBCO \cite{rmyama}. In the absence of
coherent interplane hopping \cite{raipt}, a different explanation of
this modulation would be required. However, with regard to overall
agreement with experimental observation, we do not consider the
results given in Figs.~5-7(b) to be a satisfactory description of
the measured magnetic response of YBCO materials
\cite{rmdamhhasl,rmea,rkea}. Recent experiments show few features
in the spin excitation spectrum other than a strong and remarkably
narrow peak at 41meV in the superconducting state, although there is
evidence for some broad spectral weight below this energy in the
normal state. The peak occurs only close to the wavevector ${\bf Q}$,
and does not seem to disperse significantly with ${\bf q}$ or
$\omega$, so that the band picture given above, and proposed by many
authors, would appear to be a poor candidate for a full explanation.
This is in contrast to the spin spectrum of the single-layer LSCO
material \cite{rTKF,rmahrm,rye}, where the current form of theory gives
a good account of the experimental features. In YBCO, the narrow and
non-dispersive nature of the 41meV peak suggests the existence of a
resonance, and there has been a theoretical proposal for this within
the class of tight-binding, strongly-correlated systems in two
dimensions \cite{rdz}. A full investigation of this possibility in
the present framework will be the subject of a subsequent
publication. Finally, we comment that no measurements of dynamic
susceptibility have so far been performed on BSCCO materials,
because these have proven extremely difficult to grow as high-quality
single crystals of the required size \cite{rhsa}.

\section{Phonon Anomalies}

	The coupling between the spin and lattice degrees of freedom
in the $t$-$J$ model has been investigated recently in Ref. \cite{nkf}.
These authors use the modulation of the parameters $t_{ij}$ and $J_{ij}$
by the $c$-axis oscillation of in-plane O atoms in a single-layer model
to explain the superconductive phonon anomalies in the important and
well-characterised $B$-symmetric modes of the YBCO plane, as well as
to obtain qualitative agreement with the results for other materials
classes, with the observation of the spin gap in underdoped YBCO
compounds, and with isotope effects at optimal doping. The reader is
referred to this paper for the background to this section.

	In Fig.~8(a) are shown the four $c$-axis modes of planar O
atoms in the bilayer system: at the single-layer level \cite{nkf},
there is no difference between the 340cm$^{-1}$ $B_{1g}$ mode and
the 193cm$^{-1}$ $B_{2u}$ mode, or between the 440cm$^{-1}$ $A_{1g}$
mode and the 307cm$^{-1}$ $A_{2u}$ mode. While this may be realistic
for the $B$-symmetric modes, which develop no net dipole moment, it
was found not to afford an acceptable description of either the
$A_{2u}$ mode, where there will be significant interplane charge
motion during the phonon oscillation, or the $A_{1g}$ mode, in which
charge may be transferred out of the bilayer unit to neighbouring
planes. Here we investigate the phonon anomalies in the bilayer in
the presence of finite interplane coupling and superexchange, which
as above lead to a finite interplane uniform spin order parameter,
but no interplane singlet order. There have been indications within
a phenomenological weak electron-phonon coupling model \cite{rhfk}
that interlayer hopping terms are sufficient to reproduce the phonon
anomalies measured \cite{rliv} in the infrared-active $A_{2u}$ mode.

	Following Ref. \cite{nkf}, in the bilayer it is necessary
to recast the phonon-spinon coupling vertices in terms of the
diagonalising quasiparticle operators. Each operator
$f_{k}^{\alpha}$ is a sum of 4 operators with coefficients given
by (\ref{edm}), so that each vertex is a sum of 16 terms, each
involving a contribution from both spin orientations. For the
``uniform RVB'' vertex one obtains on substitution
\begin{eqnarray}
\sum_{\sigma} \bar{\chi} f_{k \sigma}^{1(2) \dag}
f_{k \sigma}^{1(2)} & = & \mbox{$\frac{1}{2}$} \left[ C_{1}
\gamma_{1 k}^{\dag} \gamma_{1 k} + S_{1} \gamma_{1 k}^{\dag}
\gamma_{1 -k}^{\dag} \mp C_{12} \gamma_{1 k}^{\dag} \gamma_{2 k} \mp
S_{12} \gamma_{1 k}^{\dag} \gamma_{2 -k}^{\dag} \right. \label{euv}
\nonumber \\ & & \;\; + S_{1} \gamma_{1 -k} \gamma_{1 k} + C_{1}
\gamma_{1 -k} \gamma_{1 -k}^{\dag} \mp S_{12} \gamma_{1 -k}
\gamma_{2 k} \pm C_{12} \gamma_{1 -k} \gamma_{2 -k}^{\dag} \\ & &
\;\; \mp C_{12} \gamma_{2 k}^{\dag} \gamma_{1 k} \mp S_{12}
\gamma_{2 k}^{\dag} \gamma_{1 -k}^{\dag} + C_{2}
\gamma_{2 k}^{\dag} \gamma_{2 k} + S_{2} \gamma_{2 k}^{\dag}
\gamma_{2 -k}^{\dag} \nonumber \\ & & \;\;\;\; \left. \mp S_{12}
\gamma_{2 -k} \gamma_{1 k} \pm C_{12} \gamma_{2 -k}
\gamma_{1 -k}^{\dag} + S_{2} \gamma_{2 -k} \gamma_{2 k} - C_{2}
\gamma_{2 -k} \gamma_{2 -k}^{\dag} \right] ,
\nonumber
\end{eqnarray}
and for the ``singlet-RVB'' vertex
\begin{eqnarray}
(\Delta_{k}^{\ast} f_{k \uparrow}^{1(2)} f_{k \downarrow}^{1(2)}
+ {\rm {h.c.}}) & = & \mbox{$\frac{1}{2}$} \left[ - S_{1}
\gamma_{1 k} \gamma_{1 k}^{\dag} + C_{1} \gamma_{1 k} \gamma_{1 -k}
\pm S_{12} \gamma_{1 k} \gamma_{2 k}^{\dag} \mp C_{12} \gamma_{1 k}
\gamma_{2 -k} \right. \label{esv} \nonumber \\ & & \;\;\;\; + C_{1}
\gamma_{1 -k}^{\dag} \gamma_{1 k} + S_{1} \gamma_{1 -k}^{\dag}
\gamma_{1 -k}\mp C_{12} \gamma_{1 -k}^{\dag} \gamma_{2 k}^{\dag}
\mp S_{12} \gamma_{1 -k}^{\dag} \gamma_{2 -k} \\ & & \;\;\;\; \pm
S_{12} \gamma_{2 k} \gamma_{1 k}^{\dag} \mp C_{12} \gamma_{2 k}
\gamma_{1 -k} - S_{2} \gamma_{2 k} \gamma_{2 k}^{\dag} + C_{2}
\gamma_{2 k} \gamma_{2 -k} \nonumber \\ & & \;\;\;\; \mp \left.
C_{12} \gamma_{2 -k}^{\dag} \gamma_{1 k}^{\dag} \mp S_{12}
\gamma_{2 -k}^{\dag} \gamma_{1 -k} + C_{2} \gamma_{2 -k}^{\dag}
\gamma_{2 k}^{\dag} + S_{2} \gamma_{2 -k}^{\dag} \gamma_{2 -k}
\right],
\nonumber
\end{eqnarray}
where $C_{\alpha}$ and $S_{\alpha}$ denote $\cos 2 \theta_{k}
^{\alpha}$ and $\sin 2 \theta_{k}^{\alpha}$, and $C_{12}$ and
$S_{12}$ are the combinations $c_{1} c_{2} - s_{1} s_{2}$ and
$c_{1} s_{2} + s_{1} c_{2}$, where $c_{\alpha}$ and $s_{\alpha}$
denote $\cos \theta_{k}^{\alpha}$ and $\sin \theta_{k}^{\alpha}$.
The complexity contained in 2$\times$2 matrix
vertices and Nambu propagators for the $\vec{f}_{k}$ problem is
transferred solely to the 4$\times$4 matrix vertex in the
quasiparticle problem, conserving the number of degrees of freedom.

	In the even modes ($A_{1g}$ and $B_{1g}$), the motions of
the O atoms are out of phase between the planes, so that the effect
of their motion on $t$ and $J$ is the same in each plane. Defining
the phonon coordinate as positive when the oscillating O atom
moves towards the plane of the Cu atoms (Fig.~8(b)), one has in
obvious notation $u_{x}^1 = u_{y}^1 = u_{x}^2 = u_{y}^2 = u$ for
$A_{1g}$ and $u_{x}^1 = - u_{y}^1 = u_{x}^2 = - u_{y}^2 = u$ for
$B_{1g}$. Thus in both cases the effective vertex from summing the
contributions of both planes in (\ref{euv}) and (\ref{esv}) above
is to cancel all of the terms mixing the indices 1 and 2 for both
$x$ and $y$ bonds, leaving only those terms related to
intraband processes. Conversely, the odd modes $A_{2u}$ ($u_{x}^1
= u_{y}^1 = - u_{x}^2 = - u_{y}^2 = u$) and $B_{2u}$ ($u_{x}^1 =
- u_{y}^1 = - u_{x}^2 = u_{y}^2 = u$)
will have vertices coupling only to interband scattering
processes. This very simple but profound separation of the
contributions may be understood readily, in that the even modes
produce a pattern of modulations of $t$ and $J$ which is even
in the unit cell, leading only to intraband transitions, and
conversely for the pattern of modulations arising from the odd
modes. The lowest-order quasiparticle polarisation terms
contributing to the phonon self-energy may then be calculated to
deduce the frequency shift and linewidth broadening of each type of
mode. For the even modes (intraband), the contribution from normal
scattering processes will vanish in the limit $q \rightarrow 0$, and
one may follow \cite{nkf} exactly to obtain
\begin{equation}
\delta \omega = c \left( \lambda_J J \right)^2 \frac{2}{N} \sum_{k}
F_{k}^{+} \frac{1}{\omega^2 - \left( 2 E_{k}^{+} \right)^2}
\frac{\tanh \left( \frac{E_{k}^{+}}{2 T} \right)}{E_{k}^{+}} +
F_{k}^{-} \frac{1}{\omega^2 - \left( 2 E_{k}^{-} \right)^2}
\frac{\tanh \left( \frac{E_{k}^{-}}{2 T} \right)}{E_{k}^{-}} .
\label{edoe}
\end{equation}
Here $c = \mbox{$\left( \frac{3}{4 a} \right)^2$} \langle u^2
\rangle = 1.18 \times 10^{-4}$, $\langle u^2 \rangle =
\mbox{$\frac{\hbar}{2 M \omega_0}$} = ( 0.055 {\rm {\AA}} )^2$, in
which $\omega_0$ = 340cm$^{-1}$ is taken from the $B_{1g}$ phonon
frequency and $M$ is the mass of the O atom, and $E_{k}^{\pm}$ are
given by (\ref{eme}). The form factors $F_{k}^{\pm}$ are
\begin{eqnarray}
F_{k}^{\pm} & = & 2 \Delta^2 \left[ \gamma_k \left( \xi_{k} \pm
|M_k| \right) + \mbox{$\frac{3 J}{4}$} {\bar {\chi}} \eta_{k}^2
\right]^2 \;\;\;\;\;\;\;\;\;\; B_{1g} \label{effe} \nonumber \\
F_{k}^{\pm} & = & 2 \Delta^2 \eta_{k}^2 \left[ \left( \xi_{k} \pm
|M_k| \right) + \mbox{$\frac{3 J}{4}$} {\bar {\chi}} \gamma_k
\right]^2 \;\;\;\;\;\;\;\;\;\; A_{1g},
\end{eqnarray}
where $\gamma_k = \cos k_x + \cos k_y$, $\eta_k = \cos k_x - \cos
k_y$, $\Delta = \langle \Delta_{ij} \rangle$ and ${\bar {\chi}}
\equiv \langle \chi_{ij} \rangle + \frac{2 t \delta}{3 J}$. In this
calculation we have not included an ``interplane uniform'' vertex
with contributions proportional to $B_{k}^{\prime}$ and
$\chi_{k}^{\prime}$, and will discuss the effects of such terms in
detail below. The linewidth broadening can be computed from an
analogous expression for the imaginary part of the self-energy.

	For the odd modes, the interband scattering expression and
form factors become more complicated. One obtains
\begin{eqnarray}
\delta \omega & = & c \left( \lambda_J J \right)^2 \frac{1}{N}
\sum_{k} F_{k}^{n} \frac{2 \left( E_{k}^{+} - E_{k}^{-} \right)}
{\omega^2 - \left( E_{k}^{+} - E_{k}^{-} \right)^2} \left(
f(E_{k}^{-}) - f(E_{k}^{+}) \right) \label{edoo} \nonumber \\
& & \;\;\;\;\;\;\;\;\;\;\;\;\;\;\;\;\;\;\;\; + F_{k}^{a}
\frac{2 \left( E_{k}^{+} +
E_{k}^{-} \right)}{\omega^2 - \left( E_{k}^{+} + E_{k}^{-}
\right)^2} \left( 1 - f(E_{k}^{-}) - f(E_{k}^{+}) \right) ,
\end{eqnarray}
in which the form factor for normal scattering events is
\begin{eqnarray}
F_{k}^{n} & = & \left[ \sqrt{2} \Delta \gamma_k ( c_1 s_2 + s_1
c_2 ) + {\bar {\chi}} \eta_k ( c_1 c_2 - s_1 s_2 ) \right]^2
\;\;\;\;\;\;\;\;\;\; B_{2u} \label{effon} \nonumber \\ F_{k}^n
& = & \left[ \sqrt{2} \Delta \eta_k ( c_1 s_2 + s_1
c_2 ) + {\bar {\chi}} \gamma_k ( c_1 c_2 - s_1 s_2 ) \right]^2
\;\;\;\;\;\;\;\;\;\; A_{2u} ,
\end{eqnarray}
and that for anomalous scattering events is
\begin{eqnarray}
F_{k}^{a} & = & \left[ \sqrt{2} \Delta \gamma_k ( c_1 c_2 - s_1
s_2 ) + {\bar {\chi}} \eta_k ( c_1 s_2 + s_1 c_2 ) \right]^2
\;\;\;\;\;\;\;\;\;\; B_{2u} \label{effos} \nonumber \\ F_{k}^a
& = & \left[ \sqrt{2} \Delta \eta_k ( c_1 c_2 - s_1
s_2 ) + {\bar {\chi}} \gamma_k ( c_1 s_2 + s_1 c_2 ) \right]^2
\;\;\;\;\;\;\;\;\;\; A_{2u} .
\end{eqnarray}
It is straightforward to show that in the limit where the bands
become degenerate, both sets of expressions reduce to the
result expected from \cite{nkf} for the single layer.

	In Figs.~9-12 we show the phonon frequency shift $\delta
\omega$ calculated using the above formulae, and the linewidth
broadening $\delta \Gamma$, which is calculated from analogous
expressions. We compute these anomalies only for the YBCO system,
where the buckling of the O atoms in the bilayer structural unit
is uniformly directed along the $c$-axis towards the symmetry plane
of the unit cell, an arrangement which may yield constructive
combination of the vertex contributions to give a linear spin-phonon
coupling term in cases where the symmetries of the phonon distortion
and the superconducting gap are compatible. By contrast, in BSCCO
the single layer is orthorhombically buckled, into a tilting
distortion which is odd
in the unit cell \cite{rhsa}, so there is no linear coupling term
and the anomalies are expected to be negligibly small \cite{nkf},
as confirmed by experiment \cite{rsugai}.

	In Fig.~9(a) is shown $\delta \omega$ as a function of
frequency for the $g$-symmetric modes, at $q = 0$ and at a
temperature well below the singlet ordering transition $T_{RVB}$
(Fig.~4). As in the single-layer case, low-frequency $B$-symmetric
phonons will be softened by the spinon contribution, while
high-frequency modes will show hardening. The characteristic
frequency of the sign change is set by the value of the gap
$\Delta_k$ near the heavily-contributing $(\pi,0)$ points for the
intraband processes in question. Quantitatively, it is somewhat
smaller than in the single-layer solution, requiring that the
analogue of the maximally-shifted $B_{1g}$ mode in this system appear
at $\omega \simeq 0.2J$, in comparison with the experimental mode
frequency $\omega_0 \simeq 0.3 J$. The imaginary part (Fig.~10(a))
has the expected form of a peak at the frequency where the real part
changes sign, and shows that only phonon modes with frequencies close
to this value will be significantly broadened. For even, $A$-symmetric
modes one finds also the result of the single-layer system, that the
anomalies are negligible. The symmetry effects which cause the linear
contribution to be small in this case are not affected by the
introduction of interplane terms.

	For odd ($u$-symmetric) modes one sees a different feature
(Figs.~9(b),10(b)). Because these couple only to interplane scattering
processes, the characteristic frequency which determines the change
from phonon softening to hardening is set not by twice the gap
$\Delta_k$, but by twice the interband energy separation $M_k \simeq
\delta t_{\perp k}$, again at the $(\pi,0)$
points where it is largest. From the results of Ref. \cite{roka}
which were used in setting the interplane hopping parameter
$t_{\perp}^0$, and are consistent with available experimental evidence,
this separation is of order $J$, and so appears on an altogether
different energy scale. Thus all $u$-symmetric phonons in the usual
energy range will soften at the singlet ordering transition within
such a model, and the linewidth broadening will be negligible.

	In Figs.~11(a) and~12(a) are shown the temperature dependences
of $\delta \omega$ and $\delta \Gamma$ for the $g$-symmetric modes;
this is the quantity measured by experiment \cite{ralt,rprperh}, and
should be directly comparable with the results for the $B_{1g}$ and
$A_{1g}$ modes. One sees again that the results are very similar to
the single-layer case: for the $B_{1g}$ mode there is a gradual
development of the frequency shift over a range of $T$ below its onset
at $T_{RVB}$, and a peak in the linewidth broadening there, both in
good agreement with experiment \cite{ralt}. The chosen mode frequency
here is $\omega_0 = 0.2 J$, which is somewhat below the experimental
value, but is consistent with the location of the $B_{1g}$ phonon
being that for a maximal anomalous effect, and remains in a reasonable
degree of correspondence with the physical system given the nature of
the model. For the
$A_{1g}$ mode, the chosen frequency should be in the phonon
hardening regime, but we show again $\omega_0 = 0.2 J$ to emphasise
that there is very little effect, and we comment on this in detail
below. In Fig.~11(b) is the $T$ dependence of $\delta \omega$ for a
mode of frequency $\omega_0 = 0.2 J$ and $B_{2u}$ symmetry. Here
$\omega_0$ is so far from the characteristic frequency of the
changeover from softening to hardening of $u$-symmetric phonons
(Fig.~9(b)) that the development of the anomalous frequency shift is
effectively immediate at $T_{RVB}$, as seen less obviously in the
single layer model, and in good agreement with neutron scattering
studies of this mode in YBa$_2$Cu$_3$O$_7$ \cite{rhs}.
Similarly, $\delta \Gamma$ for such phonons as a result of coupling
to the spin is negligible within this model.

	We have considered only $q = 0$ phonons here, for comparison
with the results of Raman and infrared spectroscopy. For finite
in-plane wavevector ${\bf q}$, one may expect the results to be
similar to the single-layer case \cite{nkf} for both $g$- and $u$-
symmetric modes, namely that the anomalies show some features of the
Fermi surface shape, and fall to zero as ${\bf q}$ approaches the
$(\pi,0)$ and $(\pi,\pi)$ points. However, at finite $q_z$ the
intra- and interband scattering processes become mixed, so that
both mode symmetries may show both characteristic energy scales. Of
particular note is the situation at ${\bf q} = 0$, $q_z = \pi$, where
the anomalies will remain large and the contributing processes will
be completely reversed in (intra- or interband) nature. Thus
it should be possible by neutron scattering to see that the onset of
the frequency shift $\delta \omega$ for the $B_{1g}$ phonon is very
much sharper in temperature, if smaller in overall magnitude, as its
frequency will no longer be close to the relevant energy scale.
Similarly, no alterations in the linewidth of this phonon should be
detectable.

	We have shown that the bilayer system retains a good
description of the main qualitative features of the $B$-symmetric
modes of planar O in the bilayer structural unit, although we stress
again that the $g$- and $u$-symmetric modes have very different
characteristic energy scales in the coupled bilayer system.
However, the current level of treatment does
not contain the physics required to explain the experimental results
for the $A$-symmetric modes. Here we have included the interlayer
hopping only through the way in which it arises in the quasiparticle
transformation, and in this sense $t_{\perp k}$ is treated on an equal
footing with the terms $t_{k}^{\prime}$ and $t_{k}^{\prime\prime}$,
which are taken to determine $E_{k}^{\pm}$ but not to contribute
qualitatively to the phonon anomalies \cite{nkf}. This is also the
approach taken by Ref. \cite{rhfk}, but these authors chose the value
of $t_{\perp}^0$ to coincide with the phonon frequency, in order to
enhance the anomaly in their approach, and in addition used both less
restrictive symmetry constraints and an isotropic gap function
$\Delta_0$.

	We have neglected
the possibility of a contribution from modulation of the magnitude
of the parameter $t_{\perp}^0$, because in none of the phonon modes
considered do the Cu atoms of the bilayer change their relative
separation. If one wishes to include such a modulation, for example
on the grounds that additional transfer may occur through the
neighbouring O atoms (whose separation changes in $g$-symmetric
modes), it would appear as a term $\lambda_{\perp} \chi_{k}^{\prime}
\left( f_{k}^{1 \dag} f_{k}^{2} + {\rm {h.c.}} \right)$ in the
Hamiltonian. Substitution of the quasiparticle operators yields the
same type of vertex terms as in the expressions (\ref{euv}) and
(\ref{esv}), but with considerably smaller coefficient for reasonable
values of the interplane phonon coupling $\lambda_{\perp}$. Thus
the additional term would neither make a significant contribution
nor appear with a symmetry factor which combines constructively with
$A$-symmetric phonons. For this last reason, even a very strong
modulation of the coefficient $t_{\perp}^0$ would not explain the
experimental observations within the current theory. Such a
modulation may in fact be present in the $A_{2u}$ mode, where a
very strong dipole moment is developed during the oscillatory
motion, but it appears that this must be described by additional
physics, related to the strong effects of the motion on interlayer
charge transfer. Similarly, in the $A_{1g}$ mode one must also
take account of charge
transfer, which may occur not between the planes of the bilayer
but between each plane and the neighbouring BaO layers (in which
O(4) is the apical oxygen atom of planar Cu(2)). Recent
Raman-scattering experiments \cite{rkallpc} suggest that the
440cm$^{-1}$ $A_{1g}$ planar mode and the 500cm$^{-1}$ $A_{1g}$
mode of apical O behave as split levels of a single, fundamental
phonon frequency, indicating very strong coupling between in- and
out-of-plane modes of these symmetries.

	We conclude this section by noting that apart from the
change in the nature of the $u$-symmetric modes discussed in detail
above, none of the qualitative features extracted from the
single-layer model \cite{nkf} are altered in the bilayer system. In
particular, the important result that the spin gap in the low-doping
regime of the model is observable due to spin-phonon coupling
\cite{rklbjb} is
retained, a point which reflects that the current model does not
contain a difference in such properties between mono- and bilayer
oxide materials. The isotope effect arising as a result of the
coupling between the lattice and the spin sector, which is responsible
for superconductivity, is identical to that deduced from the
single-layer model, because the bilayer treatment also does not
reproduce the anomalous contributions from $A$-symmetric modes.

\section{Summary}

	In summary, we have analysed the spin properties of a
bilayer CuO$_2$ system, using the mean-field, slave-boson treatment
of the extended $t$-$J$ model. We find that the existence of
$d$-symmetric spin singlet order in the plane, favoured by the
single-layer model, acts to suppress interplane order completely
below a characteristic, finite value of the interplane coupling.
At this coupling, there is a very rapid crossover to a state of pure
interplane singlet spin order, suggesting that the two types do not
coexist. Working within the physical regime of interplane coupling,
which mandates exclusively in-plane singlet pairing, we show that the
observed Fermi surfaces of both BSCCO- and YBCO-like bilayer systems
may be reproduced with realistic extended in-plane transfer
integrals.

	For each type of system, we have calculated two microscopic
and physically observable properties related to the spin sector.
The dynamic susceptibility $\chi( {\bf q}, \omega )$, computed at
the RPA level, shows the spin fluctuation spectrum at all frequencies
and wavevectors; the commensuration and excitation properties may be
explained simply in terms of Fermi-surface scattering processes, but
the remaining discrepancies between the theory and experiment in YBCO
reveal that the
problem is significantly more interesting and complex than mean-field
physics. Anomalies in the frequency and linewidth of phonon modes in
the YBCO system show a strong contrast between even ($g$-symmetric)
modes, which couple to intraband scattering processes, and odd
($u$-symmetric) modes, which couple only to interband scattering. The
model provides a reasonable account of the $B$-symmetric modes, which
have no net electric dipole moment in each layer and so are not
accompanied by charge motion, but cannot describe the $A$-symmetric
modes, which do not satisfy this condition.

	We conclude by commenting that because in this model the
physical parameter regime remains dominated by the $d$-symmetric
in-plane singlet order of the single-layer case, the results for
observable quantities in the bilayer system differ qualitatively
from those of the single layer in only a small number of features.
We have carried out semi-quantitative calculations which may be
compared in detail with experiment. This makes it possible to quantify
the strengths and limitations of the current $t$-$J$ model formulation,
to learn which properties appear due to the bilayer nature, and to
gain some evaluation of the restrictions of the mean-field theory and
RPA approaches. Thus we may deduce
the directions in which to look for the additional physics which will
be required for a more complete description of the high-$T_c$ systems.
As an example of this process, we have assumed the existence of
coherent single-particle transfer between the planes of the bilayer,
and investigated the experimental consequences. The interlayer pair
tunnelling theory \cite{raipt} argues that such transfer cannot be
coherent, and leads to some very different physical predictions, which
remain to be computed for similarly detailed comparison. Continuing
rapid improvements in crystal quality, and in a variety of experimental
techniques, will provide crucial evidence in favour of one of these
viewpoints in the near future.

\section*{Acknowledgements}

	We are grateful to S. Maekawa for sharing results prior to
publication.
This work was supported financially by the Monbusho International
Scientific Research Program No. 05044037 and the Grant-in-Aid for
Scientific Research No. 04240103 of the Ministry of Education, Science
and Culture, Japan. B.N. wishes to acknowledge the support of the
Japan Society for the Promotion of Science.

\section*{Figure Captions}

\smallskip
Fig. 1: Schematic representation of the bilayer system, showing
transfer integrals for quasiparticle hopping and spin superexchange
interactions.

\smallskip
\noindent
Fig. 2: Variation of singlet spin ordering temperature $T_{RVB}$
with interplane hopping parameter $t_{\perp}^0$, at fixed
$J_{\perp} = 0.1 J$. The in-plane transfer integrals are  $t = 4J$,
$t^{\prime} = - \frac{1}{6} t$ and $t^{\prime \prime} =
\frac{1}{5} t$.

\smallskip
\noindent
Fig. 3: Bonding ($\circ$) and antibonding ($\times$) Fermi surfaces
for bilayer systems with doping $\delta = 0.2$, shown in the first
quadrant. (a) BSCCO-like system: parameters are
chosen, following Ref. \cite{rsg}, to give one open and one closed
Fermi surface, strong nesting of the latter (antibonding) surface, and
band degeneracy in the $\Gamma$M direction. (b) YBCO-like system:
parameters are chosen, following Refs.
\cite{roka,ra124,ra123}, to give open Fermi surfaces for both bands,
but with that for the bonding band close to ``closing'', and to
lift the degeneracy in the $\Gamma$M direction.

\smallskip
\noindent
Fig. 4: Self-consistent solutions for the singlet-RVB order parameter
$\Delta(T)$. (a) BSCCO-like system. (b) YBCO-like system.

\smallskip
\noindent
Fig. 5: Dynamic susceptibility calculated as a function of ${\bf q}$
in the $2d$ Brillouin zone at fixed frequency $\omega = 0.3 J$, low
temperature and
$q_z = 0$ (intraband processes only). In our notation, the X point
corresponds to $(\pi,0)$ and is symmetrical with Y, while M
denotes the point $(\pi,\pi)$. (a) BSCCO-like system: note
strong incommensuration arising from nesting of the antibonding
band. (b) YBCO-like system: note commensurate nature of the peak
intensity.

\smallskip
\noindent
Fig. 6: Dynamic susceptibility calculated as a function of ${\bf q}$
in the $2d$ Brillouin zone at fixed frequency $\omega = 0.3 J$, low
temperature and
$q_z = \pi$ (interband processes only). (a) BSCCO-like system: note
that intensity of interband processes is considerably higher than
in the intraband case, and that they exhibit only minor
incommensuration. (b) YBCO-like system: susceptibility remains
commensurate at all $q_z$.

\smallskip
\noindent
Fig. 7: Dynamic susceptibility calculated as a function of ${\omega}$
at fixed in-plane wavevector ${\bf q} = (\pi,\pi)$, with $q_z = 0$
($\times$), $q_z = \pi / 2$ ($\Box$) and $q_z = \pi$ ($\circ$).
(a) BSCCO-like system. The main peak is due to states pushed out of
the gap, and is absent for the $q_z = 0$ case, primarily because the
antibonding Fermi surface is everywhere smaller than $(\pi,\pi)$).
The shoulder at $\omega \sim J$ indicates the energy separation above
which there are no more states separated by $(\pi,\pi)$ in ${\bf q}$
space for the BSCCO-like dispersion. (b) YBCO-like system. The two
contributions to each peak are due to gaps with different values on
each band at the wavevector ${\bf k}$ corresponding to $(\pi,\pi)$
scattering. For the YBCO dispersion, the ``shoulder'' occurs at
$\omega = 1.2 J$.

\smallskip
\noindent
Fig. 8: (a) Atomic motions and experimentally observed frequencies
of the four phonon modes involving $c$-axis motion of in-plane O
atoms in the buckled CuO$_2$ bilayer. (b) Phonon coordinate labelling
convention.

\smallskip
\noindent
Fig. 9: Phonon frequency shift $\delta \omega$ as a function of
frequency and at $T = 0.05 T_{RVB}$ in the YBCO-like system. (a)
$B_{1g}$- ($\circ$) and $A_{1g}$-symmetric ($\times$) modes:
the frequency of the sign-change is characterised by the gap
value at the strongly-contributing $(\pi,0)$ points, because only
intraband processes are involved. (b)
$B_{2u}$- ($\circ$) and $A_{2u}$-symmetric ($\times$) modes:
because the phonon coordinate couples only to interband
processes, the frequency of the sign-change is characterised by
the value of the interplane hopping term at $(\pi,0)$ points.

\smallskip
\noindent
Fig. 10: Phonon linewidth broadening $\delta \Gamma$ at $q = 0$, as
a function of frequency and at $T = 0.05 T_{RVB}$ in the YBCO-like
system. (a) $B_{1g}$- ($\circ$) and $A_{1g}$-symmetric ($\times$)
modes. (b) $B_{2u}$- ($\circ$) and $A_{2u}$-symmetric ($\times$)
modes.

\smallskip
\noindent
Fig. 11: (a) Phonon frequency shift $\delta \omega$ as a function
of $T$ for hypothetical $B_{1g}$- ($\circ$) and $A_{1g}$-symmetric
($\times$) modes of frequency $\omega_0 = 0.2 J$, at $q = 0$ for
the YBCO-like system. Because the mode frequency is close to the
characteristic frequency of the sign-change (Fig.~9(a)), the full
frequency shift develops over a range of $T$. (b) $\delta \omega$
for $B_{2u}$- ($\circ$) and $A_{2u}$-symmetric ($\times$) modes of
the same frequency. Because the $\omega_0$ is far from the
characteristic frequency of the sign-change (Fig.~9(b)), the onset
is very abrupt.

\smallskip
\noindent
Fig. 12: (a) Phonon linewidth broadening $\delta \Gamma$ as a
function of $T$ for hypothetical $B_{1g}$- ($\circ$) and
$A_{1g}$-symmetric ($\times$) modes of frequency $\omega_0 = 0.2 J$,
at $q = 0$ for the YBCO-like system. Because $\omega_0$ is
close to the frequency of the peak (Fig.~10(a)), a significant
broadening is observed. (b) $\delta \Gamma$ for $B_{2u}$- ($\circ$)
and $A_{2u}$-symmetric ($\times$) modes the same frequency. Because
$\omega_0$ is so far from the characteristic frequency of the
sign-change (Fig.~10(b)), the imaginary part is negligible.

\end{document}